# Geant4-based technical simulation study of plastic scintillators for Positron Annihilation Lifetime Spectroscopy (PALS)

D. Boras,[a,1] D. Petschke[a] and T. Staab[a]

[a] *University Wuerzburg, Department of Chemistry, LCTM Roentgenring 11, D-97070 Wuerzburg, Germany*
*E-mail*: dominik.boras@uni-wuerzburg.de

ABSTRACT: The influence of detector setup configuration and scintillator material choice on spectrum quality in Positron Annihilation Lifetime Spectroscopy (PALS) is fundamentally acknowledged primarily by empirical observation. However, this study quantifies the effects of ultra-fast plastic scintillators (BC422Q) within a conventional collinear (180-degree) detector setup used in the laboratory using the Geant4 simulation toolkit. We examine the impact of the scintillator's dimension and geometry (truncated cone vs. cylinder) on the detection efficiency for specific gamma-ray energies (1274 keV and 511 keV) and their proportion of corrupt events such as backscattering or multiple detections. Results indicate that the selection of scintillator dimension and shape significantly enhances detection efficiency albeit with an increase in corrupt events. Furthermore, we investigate the influence on the instrument response function (IRF) of the scintillators, showing how truncated cones offer superior precision and, thus, a narrower IRF compared to cylindrical shapes.

Additionally, the effect of the sample material itself, in the case of a truncated cone as the scintillator shape, on the scintillator response was studied. It is observed that with an increasing atomic number of the sample the detection efficiency substantially decreases while the proportion of corrupt events also diminishes. Despite the sample material's influence on the energy of gamma-quanta interacting with the scintillator, no measurable impact on the IRF was detected for the chosen windows of the pulse height spectra.

This investigation encourages a profound examination of the influences on spectrum quality in PALS measurements using Geant4 as a simulation tool highlighting the critical balance between detection efficiency and corrupt event frequencies.

KEYWORDS: plastic scintillator, Geant4, PALS, simulation.

# Contents

# 1. Introduction

Positron Annihilation Lifetime Spectroscopy (PALS) is recognized as a powerful tool for the non-destructive analysis of microstructures and lattice defects within a wide range of material sciences: this is ranging from metals and alloys [1-5] over elemental and compound semiconductors [6, 7] to polymeric materials [8, 9].

The sensitivity of PALS is based on the property of the positron as a probe particle on atomic scales: upon penetration into a material, positrons undergo rapid thermalization, a process during which they lose their initial kinetic energy primarily through scattering events with electrons and, subsequently, with phonons. This energy dissipation allows the positron to reach thermal equilibrium with the lattice structure of the material. Once at thermal equilibrium, the positron diffuses through the crystal lattice, effectively scanning millions of lattice sites simultaneously. Open volume defects, such as vacancies, vacancy clusters, dislocations in single-crystal metals or semiconductors, and grain boundaries in polycrystalline metals and their alloys, exhibit a lower positron repulsion as well as a lower electron density compared to the surrounding undisturbed lattice structure. Due to the absence of positive atomic core charges, an attractive potential for positrons is generated, creating energetically favourable conditions for positron trapping. Additionally, the reduced electron density results in an extended average positron lifetime before annihilation with an electron occurs [4, 10]. Thus, by analysing the variations in the distributions of the measured lifetimes, i.e. the annihilation lifetime spectra, one can infer the presence, concentration, and nature of defects within the material [11].

Technically, the lifetime of the positron is directly measured through the time differences of the START and STOP gamma-rays accompanying its creation (22-Na, 1274keV) and annihilation with an electron (511 keV). The transformation of gamma-rays into optical photons is accomplished using scintillators. These scintillators are coupled with photomultiplier tubes (PMTs) which serve to amplify the photoelectrons released from the photocathode due to the scintillation light. This amplification results in an electrical signal that carries the essential timing and energy information.

For the PALS method, different kinds of scintillators are commonly in use: this encompasses both very efficient solid-state scintillators such as $BaF_2$ [12] and L(Y)SO [13], and plastic scintillators, which deliver very fast timing signal [14, 15]. Modern developments in ultra-fast plastic scintillators combine advantages such as rapid signal decay-times with the simplicity of single-component scintillation pulses. Additionally, compared to the solid-state variants (BaF2 and L(Y)SO), these scintillators' significantly lower atomic numbers and densities minimize the occurrences of backscattering of the gamma-rays which negatively affect the spectra quality due to distortions in the signal timing [16-18]. Hence, the use of solid-state scintillator materials typically requires an off-180° alignment [19] or Lead (Pb) shielding [20], thus, making the covered solid angle so small that better efficiency is more than nullified.

In this study, we specifically deploy the Geant4 simulation toolkit to investigate the characteristics and performance of varying plastic scintillator geometries for the method of PALS.

Geant4 [21-23], a widely utilized toolkit for simulating particle interactions with matter, has become an essential tool in the positron research community for advancing the understanding of both the physical processes and technical configurations central to the PALS method [24-27]. Leveraging Geant4's capabilities, researchers can simulate various scenarios, including positron penetration depths in different materials [24, 28], the effects of backscattered positrons annihilating in the source [29], and the optimization of detector alignments [30]. These simulations provide valuable insights that inform experimental design and interpretation.

Central to the first part of our investigation is the effect of scintillators' geometry on the overall performance: there we are particularly focusing on general detection efficiencies and the occurrences of corrupting events such as misidentifications or double detections of both START and STOP gamma rays. Additionally, the time spread of the scintillator response is studied upon varying scintillator designs by simulating a zero lifetime.

In the second part of this study, we examine the impact of the sample material on the general detection efficiency and occurrences of corrupting events as well as the impact on the scintillator' response function.

## 2. Methods

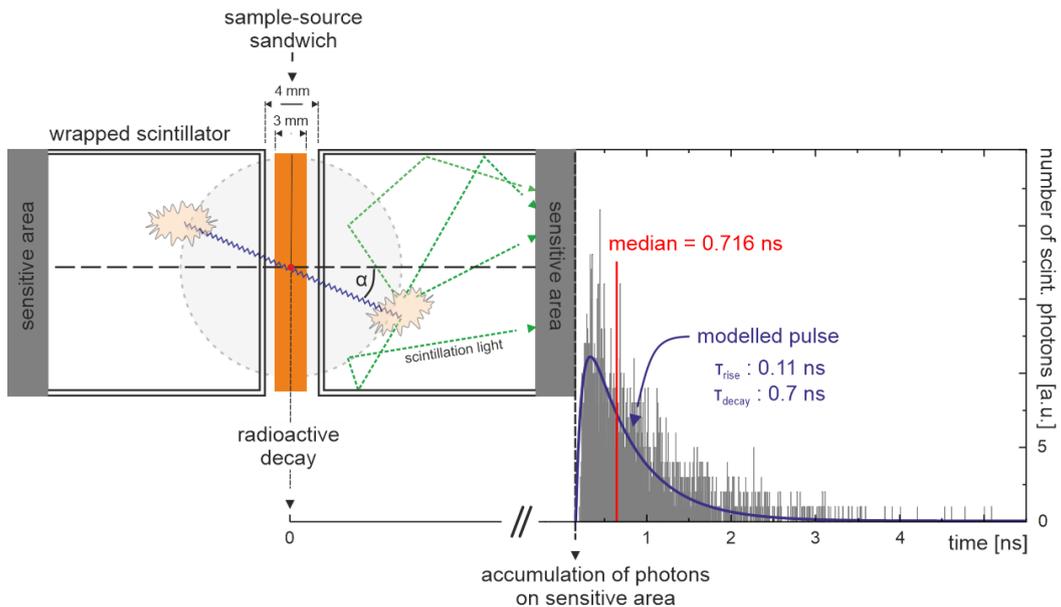

**Figure 1:** Illustration of the two-detector setup implemented for the simulation using Geant4. The scintillators are colinearly arranged (180°) with a 4 mm gap in between fitting a sample of 1.5 mm in thickness assuming a sample-source sandwich. The sample-source sandwich and the gamma-radiation source (red dot) are positioned centered to the symmetry axis of scintillator pair. The scintillators are wrapped entirely in Teflon foil ensuring a maximized count of scintillation photons reaching the sensitive area. The plot (right) shows a histogram of the arrival times of the scintillation photons accumulated at the sensitive area for a specific gamma-ray interaction event. The blue curve indicates the modelled scintillator pulse based in the parameters serving as input for the simulation as given in Table 1. The red line indicates the median of the arrival times used for the determination of the scintillator response in section 3.2.

Figure *1* shows the setup modelled in our Geant4 simulation. Our simulation disregards all components employed in a signal detection after the scintillator: Those are the photomultiplier tubes (PMTs) as we solely concentrate on the scintillators' specifications and performance properties in relation to the application of PALS. Thus, factors like optical coupling between the scintillator and the PMT through silicon oil, the PMT window material (e.g., borosilicate glass) and photocathode quantum efficiency (QE) are not considered in the simulation. However, the named components are essential for quantifying the overall performance and quality of the spectrometer but peripheral to our core objectives on the scintillator's characteristics. By following this approach, scintillation photons produced by gamma-ray interactions within the scintillator and reaching its end usually coupled to the PMT are counted at a specific sensitive area acting as a proxy for the photocathode of a hypothetical PMT covering the entire backside of the scintillator.

As illustrated in Figure *2*, scintillators with four distinct geometries are simulated employing the same plastic material BC422Q (0.5% benzophenone) [31] (Table 1). This includes three cone-shaped variants with lengths (z) of 10.0 mm, 13.57 mm and 17.15 mm (Figure *2* a-c) and a cylindrical scintillator with a length of 10 mm (Figure *2* d). The cone-shaped scintillators were specifically designed in such a way that the smallest one (10 mm) with the lowest volume (V = 7.33 cm³) features a frustum angle of 45° (Figure 2a). Conversely, the largest cone-shaped scintillator measuring 17.15 mm in length occupies the same volume (V = 12.57 cm³) as the cylindrical scintillator, thus, serves as a basis for comparison between geometric shapes. The intermediate cone-shaped scintillator, having 13.57 mm in length, has a volume of 9.95 cm³ that bridges the smallest cone-shaped scintillator (10 mm) and the cylindrical one. All scintillators have a diameter of 20 mm at the end facing the sample material, expanding to 40 mm at the end coupled to the PMT, except for the cylindrical scintillator which maintained a consistent diameter of 40 mm throughout. These specific dimensions and resulting volumes were chosen to closely reflect commonly used configurations in real-world PALS setups employing plastic scintillators [32]. To facilitate a uniform quantitative comparison across different scintillator designs, we standardized the diameter of the area facing the PMTs to 40 mm for all configurations ensuring that performance differences are attributable solely to the scintillator geometries and dimensions.

Unlike real setups, where photocathodes usually exhibit varying sensitivity across their surface, we simplified our model to assume a uniform and wavelength independent sensitivity across this complete sensitive area. This assumption implies that, theoretically, every scintillation photon reaching the sensitive area would be effectively converted into photoelectrons streamlining our comparison by removing variability in photocathode efficiency.

Wrapping the scintillators in Teflon foil has the following advantages: firstly, it avoids the loss of scintillation photons by absorption or transmission at the surface of the scintillator, secondly it forces their reflection back into the scintillator material, thus maximizing the number of scintillation photons reaching the sensitive area. Hence, in the Geant4 simulation the optical surface condition of the scintillators was configured as *polishedteflonair* (reflectivity = 100 %). This is a pre-set parameter of the internal lookup table for optical surfaces representing a mechanically polished surface covered with Teflon [21]. Since we are not interested in the absolute photon count at the sensitive area but in evaluating the relative photon yields across various scintillator designs, optimizing photon capture is aimed solely to provide an adequate statistic. This is specifically essential for an accurate determination of the scintillator pulse timing to quantify the scintillator response (section 3.2).

Since the amplification mechanism of the PMT is not involved in the simulation, the resulting scintillator pulse evidently exhibits more noise compared to the output signal of a real PMT, particularly for pulses associated with STOP events. As shown in Figure *1*, it is represented by the histogram of the arrival times of all scintillation photons collected at the sensitive areas. Consequently, employing the constant fraction (CF) principle for accurate timing determination proved to be challenging and yielded unsatisfactory outcomes. Therefore, we opted for the median as a measure, which offered an acceptable level of accuracy for our study's objectives.

To accurately simulate the scintillation process initiated by Compton scattering, evidently being the predominant interaction mechanism for plastic scintillators exposed to gamma radiation in our energy regime (~1 MeV), we employed the *G4EmStandardPhysics_option4* and *G4OpticalPhysics* packages within the Geant4 framework [21]. For the resulting pulse-height spectra (PHS), a comparison between simulation and experiment can be found in the appendix section (Appendix: Figure *15*).

Throughout this study, we used a colinear two-detector configuration (180°) for the scintillators while maintaining a horizontal gap of 4 mm between them without any vertical displacement across the varying scintillator designs. The size of the gap corresponds to a sample thickness of 1.5 mm and a distance with air of about 0.5 mm on each side to account for possible spacers through PMT caps or enclosures of the sample-source sandwich, which are typically used in PALS experiments. Moreover, the chosen sample thickness allows all positrons across the studied materials (Al, Ni, Ag, and Au) to annihilate inside the samples using 22-Na as a positron source. The sample-source sandwich and the positron source (22-Na) were placed in between both scintillators centered in relation to the symmetry axis (Figure *1*).

As our focus was on evaluating the scintillators' characteristics with exclusively gamma-ray interaction being relevant for assessing their performance and specifications, we simplified the simulation of the positron source (22-Na) to include only the gamma-rays associated with the creation and annihilation of positrons. This simplification involved a single gamma-ray with an energy of 1274 keV, designated as the START-event, for the positron's creation, and two colinear gamma-rays, each with an energy of 511 keV, designated as the STOP-event, for the positron's annihilation with an electron. To accurately simulate the radiation process, we used a uniform distribution over a sphere for the emission direction of the gamma-rays, as indicated in references [33, 34].

Positioning the sample-source sandwich in the center of a colinear scintillator arrangement evidently causes the simultaneous detection of the two 511 keV events by the opposing scintillators [17, 30, 35, 36]. This configuration, even though potentially less optimal in terms of spectra quality, is, however, commonly applied in PALS especially when utilizing plastic scintillators which inherently provide lower detection efficiency compared to solid-state scintillators such as $BaF_2$ or L(Y)SO, thus requiring longer measurement times.

The simulation was implemented to separately simulate START (a single gamma-ray: 1274 keV) and STOP (two colinear gamma-rays: 511 keV each) events. This allows for a detailed investigation of the individual effects of START and STOP events on the performance of the studied scintillator designs. To provide a comprehensive analysis, we executed 50 Mio. distinct simulations for both START and STOP events for each scintillator configuration under review. Moreover, to every START and STOP event we assigned a *source-ID, where* corresponding START and STOP events share an identical *source-ID* serving as a reference to the positron event from which they originated.

Hence, as START and STOP events individually or in association (sharing the same *source-ID*) are considered isolated from other events throughout this study, the here presented results are technically seen independent of the positron source and the sample material under investigation as they do not account for mutual influences arising from the statistical nature of radioactive decay or the individual lifetime of the positron.

**Table 1:** Listing of the relevant material properties of the studied plastic scintillators BC422Q (0.5%) serving as input for the Geant4 simulations [31].

| plastic scintillator: BC422Q (0.5% benzophenone) | |
|---|---|
| parameter | value |
| scintillation yield | 3306/MeV |
| scintillation rise-time | 110 ps |
| scintillation decay-time | 700 ps |
| reflection index | 1.58 |
| absorption length | 8.0 cm |
| density | 1.023 g/cm³ |

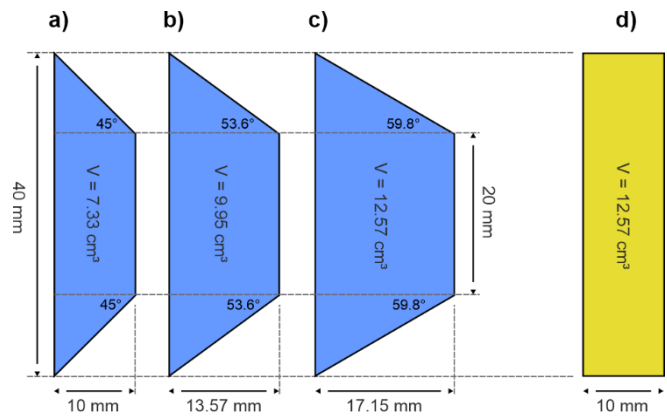

**Figure 2:** Scintillator designs: three cone-shaped configurations (a-c) with increasing length (a) 10.0 mm, (b) 13.57 mm and (c) 17.15 mm and a single cylindrical geometry (d) of length 10.0 mm are shown. The largest cone-shaped scintillator (c) occupies the same volume (V= 12.57 cm³) as the cylindrical scintillator (d), thus, serving as a basis for comparison between geometric shapes. To ensure a quantitative comparison, all scintillators have a uniform diameter of 40 mm coupled to the PMT.

## 3. Results and Discussion

### 3.1 Influence of the geometry to the frequencies of categorized event-types

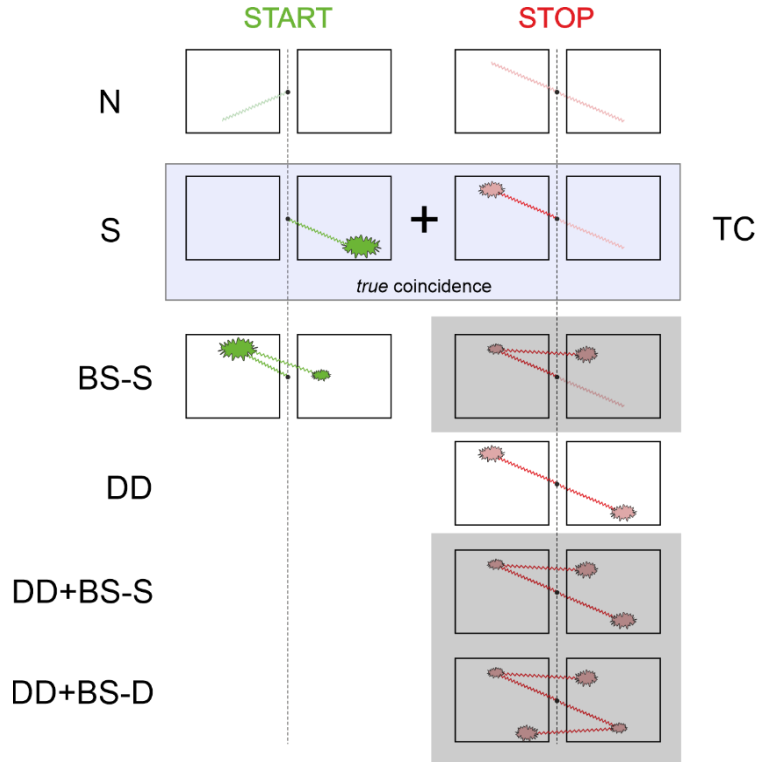

**Figure 3:** Depiction of analyzed event types: Event types within the grey-shaded boxes are excluded from this study due to their negligible frequencies compared to single detection (S), backscattered single detection (BS-S) for START and double detected (DD) STOP events. The blue box highlights the occurrence of two sequential and directly linked single START and STOP events (S) identified in opposing scintillators, categorized as true coincidence (TC).

In this section, we discuss the detection probabilities of the predominantly occurring event-types in relation to the geometry and dimensions of the scintillators (Figure *2*). The studied event-types are illustrated in Figure 3 and described in the following:

- single detection event (S): A gamma-quant directly interacts with the detector in its direction of propagation without backscattering processes involved. In the context of STOP events, this implies that exclusively one of the collinearly emitted 511 keV photons deposits energy in one of the two scintillators.
- *true* coincidence event (TC): This event-type is defined by the occurrance of two consecutive and directly associated single START and STOP events (S) detected in opposing scintillators both originating from the same *source-ID*. Furthermore, only one of the two collinearly emitted 511 keV photons interacts with the scintillator material and no backscattering occurs.
- backscattering event (BS-S): A gamma-quant initially strikes one detector in its direction of propagation, undergoes backscattering and loses energy with partial energy deposition eventually occurring in both detectors.
- double detection event (DD): This event type exclusively refers to the collinearly emitting STOP-quanta considering that both 511 keV photons strike and directly interact with the opposing detectors without backscattering processes involved.

In the analysis that follows, we included only those events in the frequency count where the number of scintillation photons accumulated at the sensitive area (as shown in Figure *1*) fall within the designated PHS windows for the respective START and STOP event types (see Appendix:

Figure *16*). For the backscattering event-type of the START-quanta (BS-S) and the double detection event-type of the STOP-quanta (DD), this condition is fulfilled, if it applies to one of the two detectors.

However, the frequencies of the backscattering-based event-types related to the STOP-quanta, specifically BS-S, DD+BS-S and DD+BS-D (depicted gray-shaded in Figure 3) are negligible in their contribution. Hence, they are not considered in this study as their backscattering primarily results in energy loss while the effective number of accumulated scintillation photons falls below the lower level (LL) of the PHS. Finally, the shown frequencies of the single detection (S) and the *true* coincidence (TC) events were normalized to encompass all generated events (50 Mio.). This includes those not striking the scintillator or undergoing interaction (N). To facilitate a more accurate comparison between START and STOP events, the single detection events (S) of the STOP-quanta were normalized to double the statistics (100 Mio. instead of 50 Mio. events) accounting for the inherent physical ratio of 2:1 between the number of emitted 511 keV and 1274 keV photons. However, the frequencies associated with corrupted events (BS-S and DD) were normalized to the sum of their detected occurrences and the number of single detection events (BS-S + S or DD + S) allowing us to quantify the contribution of corrupt events in relation to all events striking the scintillator.

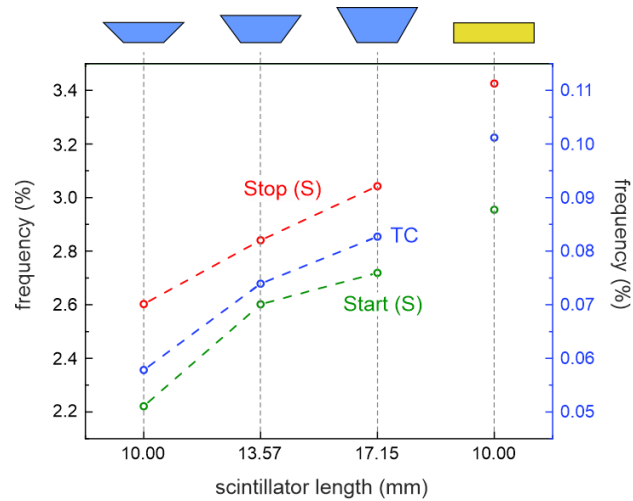

**Figure 4:** Frequencies of single detected START (green) and STOP (red) quanta separately along with *true* coincidences (TC, blue) indicating instances where two consecutive and directly linked single START and STOP events are recorded in opposing scintillators.

As can be seen in Figure 4, an increment in length (and volume) of the cone-shaped scintillator corresponds to a gradual rise in the individual frequencies of single detected START and STOP events (S). This results from the higher probability of interaction as energy deposition evidently becomes more likely for longer travel distances of the striking photons (Figure 5). The single detection (S) frequencies of the START events are consistently lower than those of the STOP events (~20%) since the probability of interaction inherently decreases for increasing photon energies [37]. However, the detection probabilities significantly rise when employing the cylindrical instead of the cone-shaped geometry. This increase attributes to the larger coverage of the solid angle (> 50-60°) providing additional volume available for gamma-ray interaction (Figure 5, yellow curve).
Consistent with expectations, this trend is also evident in the frequencies of the *true* coincidences (TC), i.e. events contributing to the resulting lifetime spectrum (Figure 4).

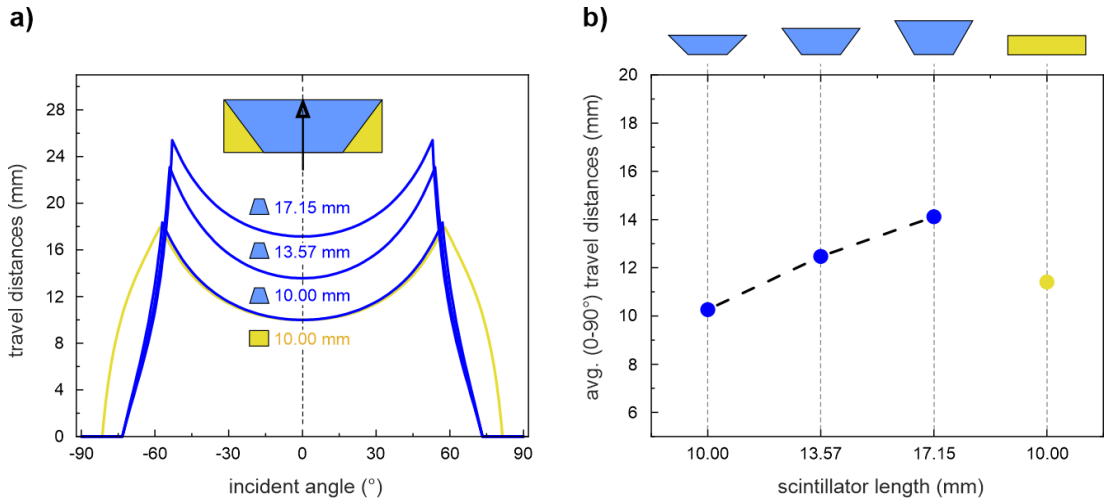

**Figure 5:** (a) Distributions of the maximum possible travel distances of the gamma rays through the scintillators across different solid angle coverages for the studied scintillator geometries and sizes with 0° indicating the axis aligned with the z-dimension that defines the length of the scintillators. (b) Average maximum possible travel distances calculated over the solid angle coverage from 0° to 90° corresponding to the distributions depicted in (a).

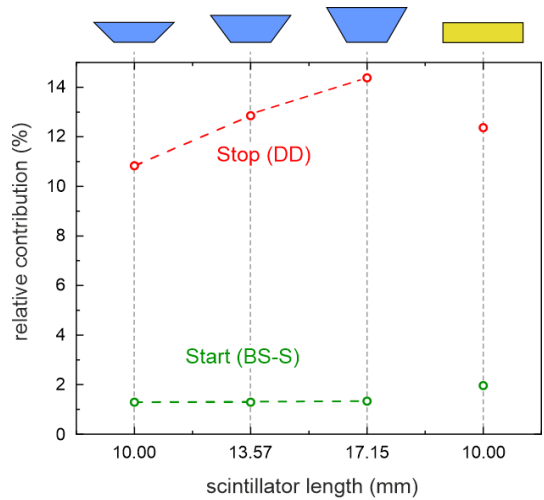

**Figure 6:** Occurrences of corrupting events relative to single detection events (S) across different scintillator geometries and dimensions. The red data points illustrate the contributions of double detected (DD) STOP events whereas green denotes the proportions of backscattered (BS-S) START events.

With regard to the occurrences of corrupted events (DD and BS-S) in relation to the single detection events (S), an ascending trend for the cone-shaped scintillator in the relative contribution of double detected STOP-events (DD) with rising length (and volume) can be noticed (Figure 6, red data points). This observation aligns with the principle that extended travel distances within the scintillator increase the probability forinteraction of the gamma-rays (cf. Figure 5 and Figure 4). In the contrary, backscattering (BS-S) of the START-quanta is less likely and the relative contribution remains almost constant below 1% over changes in length (and volume) for the cone-shaped geometry (Figure 6, green data points).

Significant differences are noted when comparing the cylindrical with the cone-shaped scintillators: The relative contribution of detected backscattering START-events (BS-S) increases notably from around 1% to about 2.5% attributed to its broader coverage of the solid angle (refer Figure 5). While this increase is relatively seen substantial (~250%), it remains minor compared to the absolute proportion of double detected (DD) STOP events considering that STOP events are inherently more prevalent due to their higher interaction probablity (Figure 4). Conversly, the relative contribution of double detected STOP-events (DD) exhibits an inverse trend for the cylindrical scintillator indicating a significant decrease to a level marginally higher than that of the cone-shaped scintillator of equivalent thickness (z=10 mm) which aligns well with the observed trend of the average maximum possible travel distances of the gamma-rays through the scintillator (Figure 5b).

To conclude, with increasing length (and volume) of the cone-shaped scintillator and the expanded solid angle coverage in the cylindrical scintillator, there is a noticeable increase in the event frequencies (S and TC) which in theory reduces the time needed for the aquisition of a lifetime spectrum. However, this benefit in efficiency is counterbalanced by a rise in the proportion of corrupting events (DD and BS-S) generally leading to a degradation of the spectra quality if not sufficiently surpressed [16, 17, 35, 36, 38].

## 3.2 Characteristics of the scintillators

In this section, the primarily relevant characteristics of the scintillators are assessed and characterized in terms of their geometry and dimensions (Figure *2*). This part of the analysis was conducted using exclusively single detection events (S) as described in the previous section.

As can be seen in Figure 7, with increasing length (and volume) of the cone-shaped scintillator, the level of maximum accumulated scintillation photons at the sensitive area (pulse-area) referring to the Compton edges of the respective START (1274 keV) and STOP (511 keV) quanta are systematically shifting towards lower values (~3200 to ~2700 photons). The observed shift is attributed to the greater absorption of scintillation photons due to their effectively longer travel distances within the scintillator as can be clearly seen in Figure 8. The narrow distributions of the travel distances indicate an effective guidance of the scintillation photons towards the photocathode in the cone-shaped scintillator. This is primarily governed by the cone's frustum angle (see also [39]). In contrast, the cylindrical scintillator displays a significantly broader distribution indicating a reduced precision in photon guidance. Evidently, the degree of broadening of the travel distance distribution directly affects the uncertainty in the time, which scintillation photons need to reach the photocathode on average, influencing the time spread (FWHM) of the scintillator's response function. As shown in Figure 9, this results in a significant increase of about 12-16 ps in the FWHM when contrasting the cylindrical (97.7 ps) with the cone-shaped scintillators (81.1 to 85.7 ps). However, the increase of about 4 ps in the time spread (81.1 to 85.7 ps) for the cone-shaped scintillators, correlated with their increasing length, is relatively minor. This increment is attributed to the modest contributions of travel distances disproportionately shifting towards lower values as scintillator length increases, as highlighted by the grey shaded area in Figure 8. Additionally, taking into account that the maximum achieved values of the pulse-areas in the PHS, corresponding to the Compton edges, are lower for the cylindrical scintillator compared to the cone-shaped scintillators, one can anticipate a notably inferior instrument response function (IRF) for the complete setup using the cylindrical scintillator. This is further compounded by the fact that the time transit spread (TTS) of the photomultiplier tubes (PMTs) is inversely related to the number of scintillation photons accumulated at the photocathode [40].

Considering our findings, employing a cylindrical geometry for the scintillator is advisable if the influence of the instrument response (FWHM) is not a critical factor and if the primary objective is achieving high efficiency and consequently reduced measurement time. This approach is particularly suitable for studying materials with long characteristic lifetimes, such as porous materials. Conversely, in scenarios where quality of the analysis takes precedence over efficiency, the cone-shaped geometry becomes preferable. This is typically the case, if the precise knowledge about the characteristic lifetimes and its contributions is essential. This is especially applicable to the study of materials having short characteristic lifetimes such as metals or semiconductors.

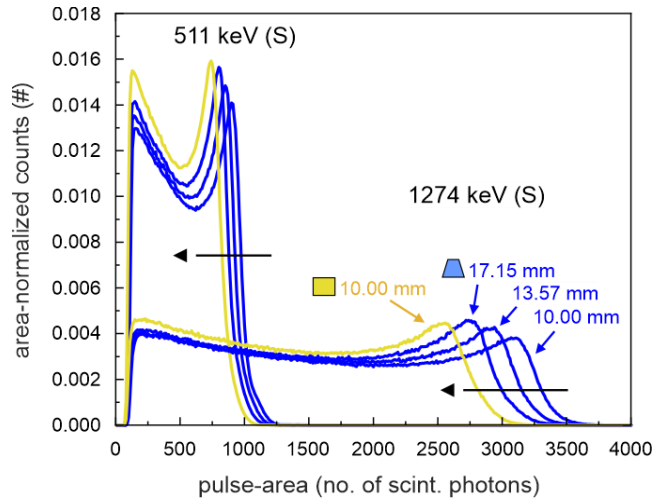

**Figure 7:** Resulting Pulse Height Spectra (PHS) based on single detected (S) START (1274 keV) and STOP (511 keV) gamma-rays for the studied scintillator geometries and dimensions. The horizontal axis shows the number of scintillation photons accumulated at the sensitive area (photocathode), which is linked to the pulse area. The arrows highlight the shift towards lower pulse areas as the length of the cone-shaped scintillator increases or the cylindrical geometry is used.

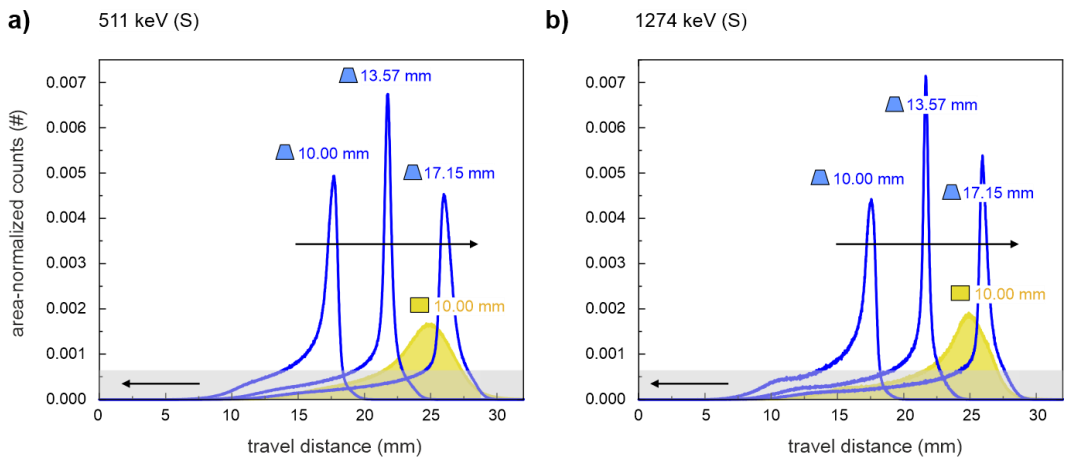

**Figure 8:** Distributions of the travel distances of scintillation photons reaching the sensitive area (photocathode) for single detected (S) START (a) and STOP (b) events for the studied scintillator geometries and dimensions. The arrows indicate the trend of the modest contributions of travel distances (grey shaded area, <0.001) disproportionately shifting towards lower values as scintillator length increases.

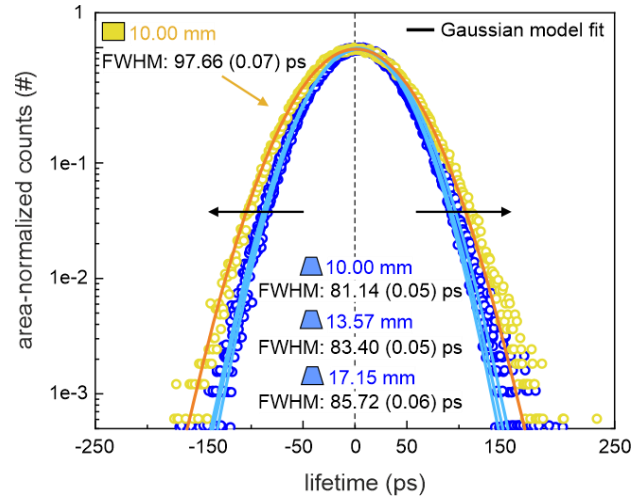

**Figure 9:** Time spread distribution in the response functions of scintillators across the various geometries and dimensions focusing solely on *true* coincidences (TC) under the simulation of a zero lifetime (0.0 ps) scenario.

## 3.3 Influence of the sample material

In this section, we examine how different sample materials (Al, Ni, Ag, and Au) affect detection probability (refer to section 3.1), the contribution of corrupt events (DD, BS-S) and the time spread (FWHM) of the scintillator response (refer to section 3.2) exemplarly in a cone-shaped scintillator with a thickness of 10 mm (Figure *2* a). The thickness of each sample material is consistently maintained at 1.5 mm, fitting in the 4 mm gap between the two facing scintillators (Figure *1*). This allows for a direct comparison with the results discussed in the previous sections. Furthermore, the lateral dimensions (16 mm diameter) and shape of the sample material were chosen to to cover most of the surface of the scintillator.

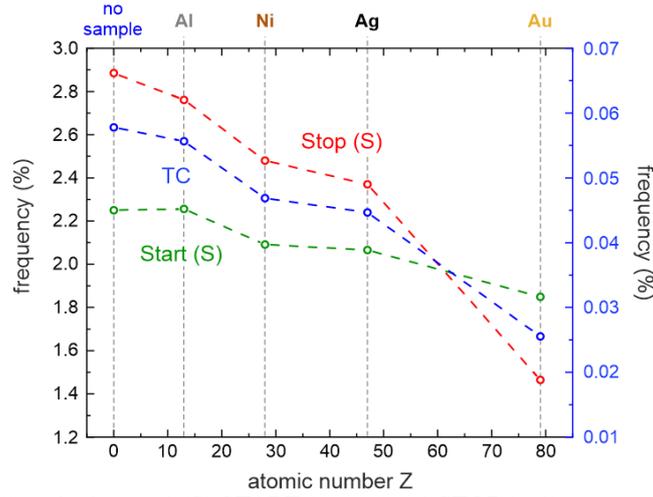

**Figure 10:** Frequencies of single detected (S) START (green) and STOP (red) quanta as well as *true* coincidences (TC, blue) for the cone-shaped scintillator with 10 mm length. We assumed sample materials of 1.5 mm thickness with different atomic numbers (Z): Al (13), Ni (28), Ag (47) and Au (79).

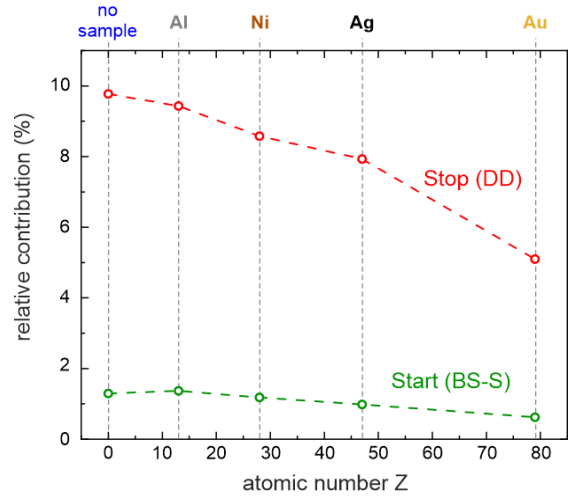

**Figure 11:** Occurrences of corrupting events in relation to single detection events (S) for the cone-shaped scintillator with 10 mm length considering sample materials of 1.5 mm thickness with different atomic numbers (Z): Al (13), Ni (28), Ag (47) and Au (79). The red data points illustrate the contributions of double detected (DD) STOP events whereas green data points denote the proportions of backscattered (BS-S) START events.

With the increasing atomic number (Z) of the materials the detection probabilities for the individual single START and STOP events (S) as well as for *true* coincidences (TC) diminish progressively (see Figure 10). This trend can be attributed to the fundamental physics of interaction processes between gamma-rays and the sample material. As the atomic number increases, material density in general also increases leading to a more pronounced attenuation of the gamma-ray energy upon interacting with the material, primarily due to Compton scattering as prevalent in this energy regime [37, 41]. As a result, a greater contribution of accumulated scintillation photons at the sensitive area does not fall within the designated windows of the PHS. Consequently, those events do not contribute to the frequency counts. Figure 10 shows that the theoretical measurement time doubles as the frequency of *true* coincidences (TC) is halved when measuring Gold (Z=79) compared to Aluminium (Z=13). Simultaneously, the relative contributions of corrupting events (DD and BS-S) decreases as the atomic number increases as shown in Figure 11. Therefore, when measuring Gold (Au) in comparison to Aluminum (Al), there is also a halving in the occurrences of double detected STOP (DD) and backscattered (BS-S) START events which in reality would lead to a higher spectra quality [16, 30, 35, 39]. This is again attributed to the progressive attenuation of the photon energy by the sample material as the atomic number increases. As a result, gamma-rays passing the sample material and striking the scintillator produce a lower number of scintillation photons accumulated at the sensitive area (pulse-area) which more frequently fall below the LL of the designated PHS windows. This effect is exemplarily depicted in Figure 12 for Aluminum (Z=13) and Nickel (Z=28). Frequencies at lower pulse-areas outside the PHS windows disproportionately rise, while there is an almost uniform decrease within the PHS windows for both START and STOP events.

Although changes in the atomic number impact the energy of gamma-rays interacting with the scintillator, the travel distance distribution of the scintillation photons remains almost unaffected by it (Figure 13). Consequently, variations in the sample material do not affect the time spread (FWHM) of the scintillator's response, as shown in Figure 14.

In summary, there is a noticeable drop in the detection efficiency with increasing atomic number of the sample material. Fortunately, this is accompanied by a simultaneous decrease in the relative frequency of corrupting events (DD and BS-S) while the time spread (FWHM) of the scintillator's response remains consistently unaffected. However, when considering other materials, incorporating PMTs may causes in differences in the IRF.

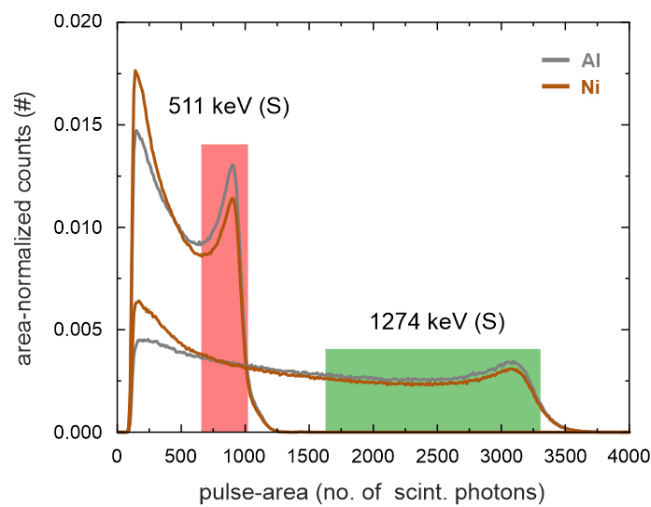

**Figure 12:** Resulting Pulse Height Spectra (PHS) for the cone-shaped scintillator with 10 mm length for the sample materials Al (13) and Ni (28) in comparison. The data are based on single detected (S) START (1274 keV) and STOP (511 keV) events. The horizontal axis shows the number of scintillation photons accumulated at the sensitive area (photocathode) linked to the pulse area.

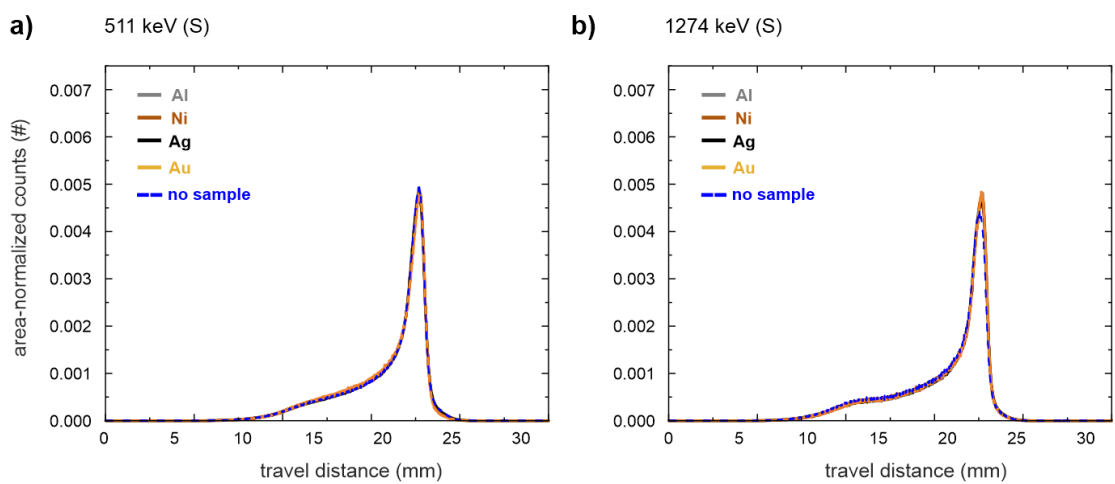

**Figure 13:** Distributions of the travel distances of scintillation photons reaching the sensitive area (photocathode) for the cone-shaped scintillator with 10 mm length considering various sample materials (Al (13), Ni (28), Ag (47)

and Au (79); thickness of 1.5 mm) between the source of radiation and the scintillators. The data are based on single detected (S) START (a) and STOP (b) events.

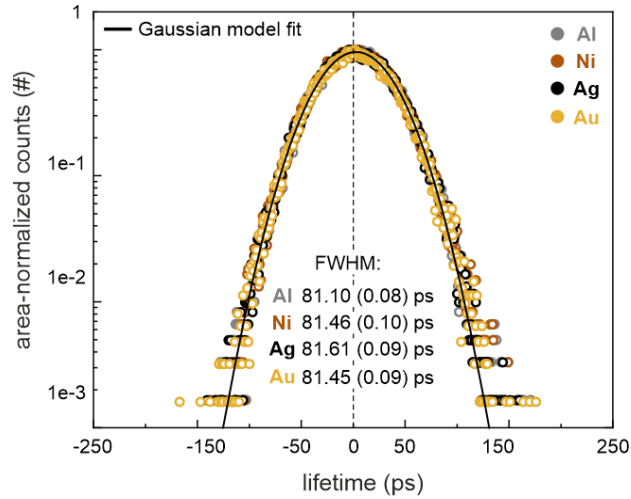

**Figure 14:** Time spread distribution in the response functions of the cone-shaped scintillator with 10 mm length considering various sample materials (Al (13), Ni (28), Ag (47) and Au (79)); thickness of 1.5 mm between the source of radiation and the scintillators. The data are solely based on *true* coincidences (TC) under the simulation of a zero lifetime (0.0 ps) scenario.

# 4. Conclusion

In this study, we systematically investigated the impact of scintillator geometry and dimensions, as well as the sample material composition, on detection probabilities with respect to efficiency and corrupting events such as backscattering and double detections, as well as the response characteristics of scintillation detectors.

Our study demonstrates that the geometry of the scintillator plays a crucial role in measurement efficiency and the resulting quality of the lifetime spectrum. Cone-shaped scintillators, with their tailored geometry, exhibit superior performance in minimizing corrupting events like backscattering (BS-S) and in maintaining a narrower time spread (FWHM) in the scintillator's response function. In contrast, cylindrical scintillators, while offering increased detection probabilities and reduced measurement times, suffer from a higher proportion of backscattering events (BS-S) and a broader time spread (FWHM). However, the cylindrical geometry, given the same volume, shows a significantly lower detection probability for corrupting events such as double detections (DD). This leads to a situation where, despite the higher detection probability of the cylindrical scintillator, a smaller proportion of 1275 keV gamma quanta are shifted in their pulse amplitude due to double detection of 511 keV gamma quanta. Thus, the choice of this geometry could significantly impair spectral quality, potentially opposing the intended narrowing of the IRF.

Furthermore, the influence of sample material, characterized by different atomic numbers (Al, Ni, Ag, and Au), was elucidated. Materials with higher atomic numbers, such as gold (Au), exhibit a significant reduction in detection efficiency and corrupting events. This reduction is attributed to the attenuation of gamma-ray energy, primarily through Compton scattering processes. This process counteracts the increase in source contribution in the spectrum from high-Z materials [29]. That is, while the intensity of the source contribution is increased due to positron scattering within the sample itself, the spectral quality improves simultaneously because corrupting events involving gamma quanta decrease with higher nuclear charge numbers.

These observations are crucial for optimizing PALS setups, particularly in selecting appropriate scintillator geometries. Additionally, the nature of the sample material under investigation should be taken into account.

In essence, this study provides valuable insights into the interplay between the characteristics of plastic scintillators and different sample materials. These insights should guide the design and application of PALS systems. For high-quality spectral analysis, particularly in the context of materials with short characteristic lifetimes such as metals or semiconductors, the choice of a cone-shaped scintillator is recommended. Conversely, for studies prioritizing detection efficiency when materials with longer characteristic lifetimes are the target, a cylindrical scintillator could be advantageous despite its noted limitations.

This work not only advances our understanding of scintillator behavior in PALS systems but also provides a solid foundation for future research.

**Author contributions**

**Dominik Boras:** Conceptualization, Methodology, Software, Formal analysis, Writing - Original Draft, Visualization
**Danny Petschke:** Conceptualization, Writing - Original Draft, Visualization, Supervision
**Torsten E. M. Staab:** Supervision, Writing - Review & Editing

**Acknowledgement**

This research did not receive any specific grant from funding agencies in the public, commercial, or not-for-profit sectors.

# 5. Appendix

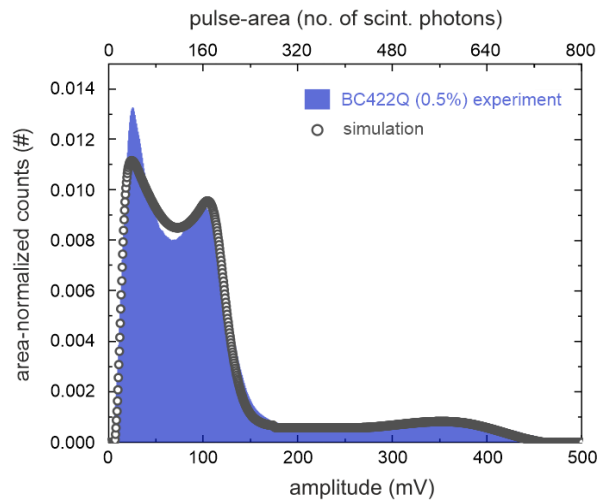

**Figure *15*:** Comparison between experimental and simulated Pulse Height Spectra (PHS) using cone-shaped plastic scintillators of type BC422Q (containing 0.5% benzophenone). The experimental setup, as described in [32], was employed, and data acquisition was facilitated using DDRS4PALS software [16, 42]. The scintillators used had dimensions with a diameter of 40 mm at the PMT interface, a length of 27.9 mm, and a diameter of 19 mm facing the sample. In the simulation, both silicone grease and a PMT entrance window made of silicate glass were incorporated, similar to the approach in the referenced publication [39]. Furthermore, a wavelength-dependent QE was included to the sensitive volume. The simulation replicated the same scintillator dimensions as those used in the experimental setup.

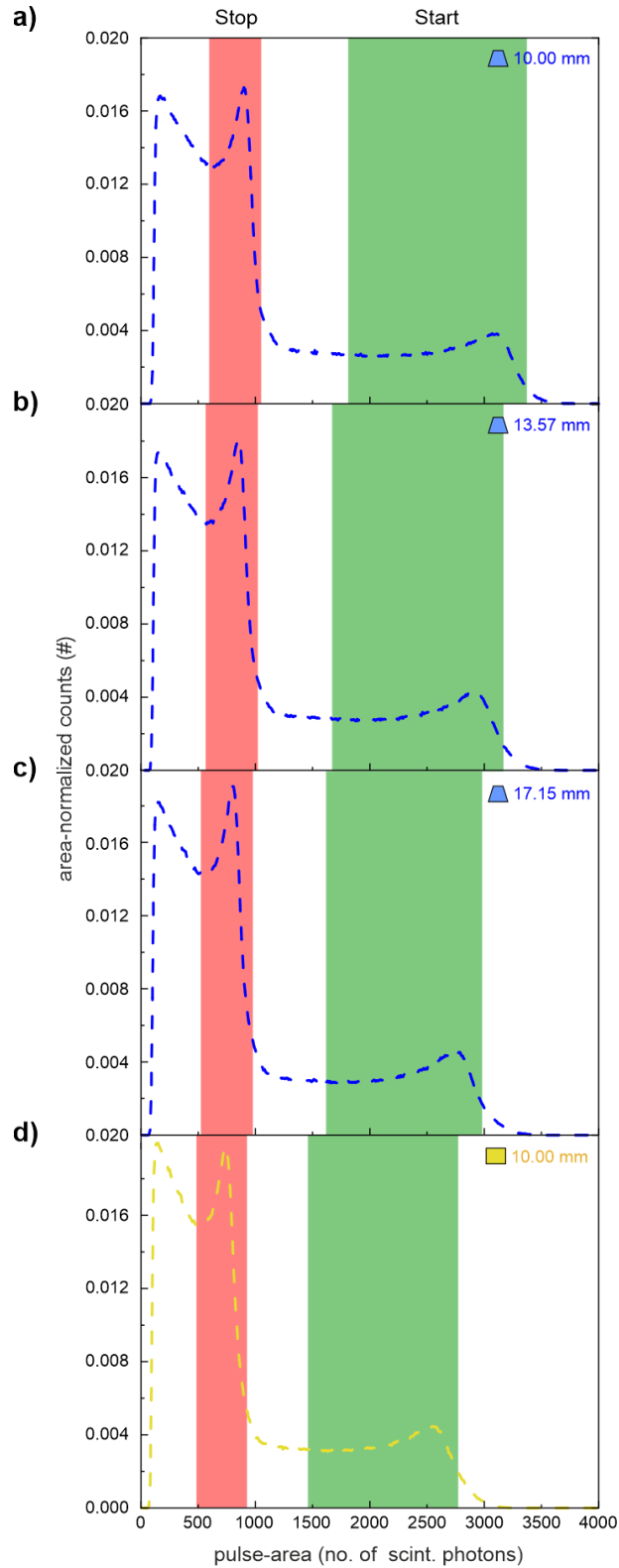

**Figure *16*:** Comparison of the Pulse Height Spectra (PHS) for the studied scintillator geometries and dimensions exclusively obtained from the single detected (S) START (1274 keV) and STOP (511 keV) gamma-rays. Areas shaded in green and red indicate the PHS windows used for counting the respective events as START and STOP to the frequencies. To counteract spectral shifts induced by the varying scintillator's geometry, detection windows were systematically adjusted. This adjustment ensures that the detection windows scale with the altered PHS and consistently encapsulate the Compton edges. Panel (a) presents the PHS for a truncated cone with a length of 10mm, panel (b) for a length of 13.57mm, panel (c)

for a length of 17.15mm and panel (d) illustrates the PHS for a cylindrical scintillator for a length of 10 mm. The horizontal axis shows the number of scintillation photons accumulated at the sensitive area (photocathode) linked to the pulse area. All PHS were normalized to the area.